\documentstyle[multicol,aps,epsfig]{revtex} 

\tighten       

\begin{document}
\draft

\title{Sequential  tunneling in doped
superlattices:\\ Fingerprints of impurity bands and photon-assisted tunneling}
\author{Andreas Wacker and Antti-Pekka Jauho}
\address{Mikroelektronik Centret,
Danmarks Tekniske Universitet, DK-2800 Lyngby, Denmark}
\author{Stefan Zeuner\cite{byline} and S.~James Allen}
\address{Center for
Terahertz Science and Technology, University of California
at Santa Barbara, Santa Barbara, California 93106}

\date{To appear in Physical Review B, 15. November 1997}
\maketitle

\begin{abstract}
We report a combined theoretical and experimental study of 
electrical transport in weakly-coupled doped superlattices.
Our calculations exhibit  negative 
differential conductivity at sufficiently high electric fields 
for all dopings. In low-doped samples the presence of impurity 
bands modifies the current-voltage characteristics substantially 
and we find two different current peaks whose relative height 
is changing with the electron temperature. 
These findings can explain the observation
of different peaks in the current-voltage characteristics
with and without external THz irradiation in low-doped samples.
From our  microscopic transport model we
obtain quantitative agreement with the experimental 
current-voltage characteristics without using any fitting parameters.
Both our experimental data and our theory show that 
absolute negative conductance persists over a wide
range of frequencies of the free-electron laser source.
\end{abstract}
\begin{multicols}{2}
\narrowtext

\section{Introduction}
Perpendicular charge transport in biased superlattices 
is dominated by resonances due to the alignment
of energy levels in different wells. These resonances yield distinct
peaks in the current-voltage characteristics \cite{ESA74,CAP86}
associated with negative differential conductivity (NDC) at fields
above the peak. The instability associated with NDC causes  the 
formation of electric field domains\cite{KWO95} as well as self-sustained 
oscillations in such structures\cite{KAS97}.
While for strongly-coupled superlattices the electronic minibands 
dominate the electrical transport\cite{ESA70}, in weakly-coupled
superlattices the transport is due to sequential tunneling from
one well to the next. (For a discussion of the appropriate regimes see 
Refs.~\cite{LAI93,WAC97}.)
This situation has been already regarded in Ref.~\cite{KAZ72}
for tunneling between the lowest level and excited levels in the
adjacent well. There the current is driven by the different occupation
of the two levels and a maximum of the current occurs when the
different level are aligned. Tunneling between
equivalent levels at low fields is slightly more complicated, as 
alignment occurs at zero field, where, of course, the current 
vanishes. The key point is the treatment of broadening 
of the states due to scattering which essentially
determines the transport. This idea has been exploited 
to determine  scattering rates by studying the transport 
between two quantum wells\cite{ZHE93,MUR95,TUR96}.
In the experiments \cite{MUR95,TUR96} impurity scattering was diminished by 
the use of remote doping which enabled to study electron-electron scattering 
rates. In contrast to this we focus on doped superlattices 
in the present paper. There impurity scattering at the ionized donors 
is an important scattering process  whose impact we
will examine in the following. 
In a previous study \cite{WAC97b} 
a heavily doped sample was investigated and good agreement 
with experimental data was found.
Here we perform a systematic study of the low-field transport
in such structures for different doping densities.
We find that the formation of impurity bands\cite{GOL88,SER89}  for 
low-doped samples causes 
a strong temperature dependence of the current-field relation
which may display a double-peak structure at low fields.

If the superlattice is subjected to an external  
microwave field, photon-assisted tunneling (PAT) is possible 
where replica of the resonances are observed at biases which 
differ from alignment conditions \cite{GUI93,KEA95a,INA96} by integer
multiples of the photon energy.
For certain field strengths of the irradiation field
absolute negative conductance has been observed 
experimentally\cite{KEA95b,ZEU96}. 
The main features of these experiments could be described 
qualitatively\cite{KEA95b,ZEU96,PLA97}
within the standard theory of photon-assisted tunneling\cite{TIE63,TUC85}
but modifications due to  photon side-bands
from a single quantum well\cite{WAG96}  have also been suggested to 
explain the experimental findings.
Here we present additional experimental data and show that
full {\em quantitative} agreement between theory and experiment
can be found by a combination of a microscopic transport model
with the standard theory of photon-assisted tunneling.
This comparison strongly supports our claim that
a microscopic treatment of impurity scattering is necessary for a 
full understanding of transport in low-doped superlattices.

The paper is organized as follows:
Our transport model is presented in Section II.
In order to understand the generic behavior we give
a phenomenological approximation where many features can be 
seen analytically in Section III.
The calculated results for different doping densities are presented
in Sections IV and V using different screening models, respectively.
Our calculations are compared with two different experiments
concerning a highly-doped and a low-doped
sample in  section VI. In section VII we consider transport
under external irradiation.
Finally, we will discuss the general significance of our results.

\section{The model}
We consider weakly-coupled semiconductor quantum wells of period $d$. 
Then the electrons are essentially localized in the wells and
a reasonable basis set of wave functions is given by
a product of Wannier functions $\Psi^{\nu}(z-jd)$, 
which are maximally localized\cite{KOH59} in well $j$, and plane waves 
$e^{i{\bf k}\cdot {\bf r}}$.
The $z$ direction is defined to be the growth direction and
${\bf k},{\bf r}$ are vectors within the $(x,y)$ plane.
$\nu$ denotes the subband within the well. Here we restrict ourselves
to the lowest level and omit the index $\nu$ in the following and
the energy of the lowest level is used as a reference point.

Regarding only next-neighbor coupling $T_1$ we have  the following 
Hamiltonian
($F$ is the electric field, and $e<0$ is the charge of the electron): 
\begin{eqnarray}
\hat{H}&=&\sum_{j,{\bf k}}\left[
(E_k-jeFd)a_j^{\dag}({\bf k})a_j({\bf k})\right.\nonumber \\
&&\phantom{\sum_{j,{\bf k}}[}\left. +T_1 a_{j+1}^{\dag}({\bf k})a_j({\bf k})
+T_1 a_{j}^{\dag}({\bf k})a_{j+1}({\bf k})\right]\nonumber \\
&&+\hat{H}^{\rm scatt}
\label{Eqham}
\end{eqnarray}
with the in-plane kinetic energy $E_k=\hbar^2k^2/(2m_w)$, 
where $m_w$ is  the effective mass  in the well.
$a_{j}$ and $a_{j}^{\dag}$ are the annihilation and creation operators
of electrons in well $j$, respectively. $\hat{H}^{\rm scatt}$ denotes
the contribution due to scattering which is not ${\bf k}$-conserving.

Within the lowest order in the coupling $T_1$
the current  density from the lowest level in well
$j$ to the lowest level in well $j+1$ is given by \cite{MAH90}
\begin{eqnarray}
J_{j\to j+1}=\frac{2e}{A}\sum_{{\bf k}}
|T_{1}|^2
\int_{-\infty}^{\infty} \frac{{\rm d}E}{2\pi \hbar}A_{j+1}({\bf k},E+eFd)\nonumber \\
\times A_j({\bf k},E)\left[n_F(E-\mu_j)-n_F(E+eFd-\mu_{j+1})\right]
\label{EqJ}\, .
\end{eqnarray}
Here $e$ is the electron charge, $A$ is the sample area, $\mu_j$ is 
the local chemical potential in well $j$ measured with respect to the energy 
of the lowest level. 
$n_F({\cal E})=1/[1+\exp({\cal E}/k_BT_e)]$,
and $T_e$ is the electron temperature.
$Fd$ denotes the voltage drop per period $d$.
The spectral function $A({\bf k},E)$ is calculated for a given
intrawell scattering $\hat{H}^{\rm scatt}$
via the retarded self-energy $\Sigma^{\rm ret}({\bf k},E)$:
\begin{equation}
A({\bf k},E)=\frac{-2 {\rm Im}\{\Sigma^{\rm ret}({\bf k},E)\}}
{\left(E-E_k-{\rm Re}\{\Sigma^{\rm ret}\}\right)^2+
\left({\rm Im}\{\Sigma^{\rm ret}\}\right)^2}\, .
\end{equation}
$\mu_j$ is related to the electron density $n_j$ in well $j$ via
the relation 
\begin{equation}
n_j=\int_{-\infty}^{\infty} {\rm d}E\,\rho_j(E)n_F(E-\mu_j)
\label{Eqdichte}
\end{equation}
with the density of states
\begin{equation}
\rho_j(E)=\frac{2}{2\pi A}
\sum_{{\bf k}}A_{j}({\bf k},E)\label{EqDOS}
\end{equation}
where the factor 2 reflects the spin degeneracy.

In our microscopic calculation we proceed as follows:
First we determine the coupling  $T_1$  as well as the Wannier
functions $\Psi(z-jd)$ for the given superlattice parameters
(see Appendix \ref{ApT1}). 
Then we calculate the self-energy $\Sigma^{\rm ret}({\bf k},E)$
for impurity scattering using the self-consistent single-site 
approximation shown  in Fig.~\ref{Figssa}.
The respective formulas are given in appendix \ref{Apsigma}.
The matrix element for impurity scattering
is calculated from the Coulomb potential of the individual 
ionized donors. Screening is treated in two different approaches,
the random-phase approximation (RPA) 
for a free-electron gas and the Thomas-Fermi approximation (TF)
using the actual density of states at the Fermi level
(see appendix \ref{Apscreen}).
Using the calculated spectral functions $A({\bf k},E)$ the
chemical potential is determined by setting $n_j=N_D$ in
Eq.~(\ref{Eqdichte}), where $N_D$ is the doping density per period.
Finally the current is calculated from Eq.~(\ref{EqJ}).
Note that all quantities used in the calculation 
are defined by the sample parameters
and no fitting parameters are used.

\section{Phenomenological description}
In this section we want to provide some insight into the
question how  scattering effects the transport.
Using a constant self-energy we derive some simple expressions 
for the current-field relation which will help to understand 
the full calculations presented in subsequent sections.

As mentioned in the introduction the level broadening
essentially determines the transport in
the sequential limit. This can be easily seen in the limit
of vanishing scattering. Then the spectral functions
become $\delta$-functions, $A({\bf k},E)=2\pi \delta(E-E_k)$.
In this case the current vanishes for $eFd\neq 0$. (In addition
further resonances may occur at finite fields, when the lowest level is
aligned with higher levels in the neighboring well, which are not 
considered here). Assuming $eFd\ge 0$
we rewrite Eq.~(\ref{EqJ}) as follows
\begin{eqnarray}
J_{j\to j+1}&=&e\frac{T_1^2}{\hbar}\int_{-\infty}^{\infty} {\rm d}E 
\langle A_{j+1}\rangle(E,F)\rho_j(E)\nonumber \\
&&\times \left[n_F(E-\mu_j)-n_F(E+eFd-\mu_{j+1})\right]
\end{eqnarray}
with 
\begin{equation}
\langle A_{j+1}\rangle(E,F)=
\frac{\int_0^{\infty} {\rm d}E_k \,
A_j({\bf k},E) A_{j+1}({\bf k},E+eFd)}
{\int_0^{\infty} {\rm d}E_k\, A_j({\bf k},E)}
\label{Eqfhilf}
\end{equation}
where we used Eq.~(\ref{EqDOS}) and performed the continuum limit.
Now let us assume a constant self-energy 
$\Sigma^{\rm ret}({\bf k},E)=-i\Gamma/2$ in Eq.~(\ref{Eqfhilf})
for the sake of simplicity. Then the
spectral functions become Lorentzians 
$A({\bf k},E)=\Gamma/[(E-E_k)^2+\Gamma^2/4]$.
Extending the lower integration to $-\infty$ we obtain
\begin{equation}
\langle A_{j+1}\rangle=\frac{2\Gamma}{(eFd)^2+\Gamma^2}
\end{equation}
which only depends on $F$.
Note that this simple model with a constant self-energy cannot be used
in  the calculation of the density of states (\ref{EqDOS}) as
the integral for the electron density (\ref{Eqdichte}) 
diverges in this case. Therefore we use
the free-electron density of states $\rho_j(E)=\rho_0\Theta(E)$
(with $\rho_0=m/(\pi\hbar^2)$)
and obtain for equal chemical potentials $\mu_j=\mu_{j+1}=\mu$:
\begin{equation}
J(F)=e\rho_0\frac{T_1^2}{\hbar}\frac{2\Gamma}{(eFd)^2+\Gamma^2}
\int_0^{eFd} {\rm d} E\, n_F(E-\mu)\, .
\label{EqJgamma}
\end{equation}
For low electron temperature and voltage drop ($k_BT_e,eFd\ll \mu$)
we find
\begin{equation}
J(F)=e\rho_0\frac{T_1^2}{\hbar}\frac{2\Gamma eFd}{(eFd)^2+\Gamma^2}
\label{EqJgamma1}
\end{equation}
Thus, we recover an ohmic behavior for low fields $eFd\ll \Gamma$, a
maximum of $J(F)$ at  $eFd=\Gamma$, and negative differential conductivity
for $eFd>\Gamma$. 
Eq.~(\ref{EqJgamma1}) has been essentially used in
Refs.~\cite{MUR95,TUR96} for the determination of 
scattering rates $\Gamma/\hbar$ from tunneling between two two-dimensional 
electron gases. Similar models using a phenomenological broadening 
$\Gamma$ have been applied to the sequential tunneling in
superlattices in Refs.~\cite{PLA97,MIL94}.
The current at the maximum is given by
\begin{equation}
J_{\rm max}=J\left(\frac{\Gamma}{ed}\right)=
e\rho_0 \frac{T_1^2}{\hbar}\label{EqJmaxgamma}
\end{equation}
which is independent of doping, scattering, and temperature in the limit
of $\mu\gg eFd,k_BT_e$ considered here.

If $k_BT_e$ becomes of the order of $\mu$ the factor 
$\int_0^{eFd} {\rm d} E\, n_F(E-\mu)$
in Eq.~(\ref{EqJgamma}) is smaller than $eFd$ and we obtain a decrease
of the current with temperature.
Here we have to take into account the temperature dependence
of the chemical potential $\mu$. From Eq.~(\ref{Eqdichte}) we find
$1+\exp(\mu/k_BT_e)=\exp(n/\rho_0k_BT_e)$. This gives a zero field
conductivity
\begin{equation}
\frac{dJ}{dF}_{|F=0}=\frac{2 e^2 \rho_0}{\hbar} \frac{dT_1^2}{\Gamma}
\left[1-\exp\left(-\frac{n}{\rho_0k_BT_e}\right)\right]
\label{Eqcond}
\end{equation}
which is almost constant for $k_BT_e< n/\rho_0$ and drops as $1/k_BT_e$ for
large temperatures as observed experimentally in
Ref.~\cite{BRO90a}.
For completeness, we give the result in the high temperature limit
($k_BT_e\gg eFd,n/\rho_0$)
\begin{equation}
J(F)=\frac{en}{k_BT_e}\frac{T_1^2}{\hbar}\frac{2\Gamma eFd}{(eFd)^2+\Gamma^2}
\label{EqJgamma2}
\end{equation}
which follows directly from Eq.~(\ref{EqJgamma}).
It is interesting to note,
that Eq.~(\ref{EqJgamma1}) is identical
to the current-field relation calculated for  miniband
conduction \cite{LEB70} using a constant scattering time $\hbar/\Gamma$
for $\mu>2T_1$ and $k_BT_e=0$.
Equivalently, Eq.~(\ref{EqJgamma2}) has been obtained from miniband
transport in the limit $k_BT_e\gg 2T_1,n/\rho_0$ as well\cite{SHI75}.
This shows that the models of sequential tunneling and miniband conduction
give the same results provided either the electron temperature or the
electron density are large.

\section{Results for RPA screening}
As a model system we choose an Al$_{0.3}$Ga$_{0.7}$As-GaAs superlattice
with barrier width $b=10$ nm  and well width $w=10$ nm.
We use the conduction band offset 240 meV and the effective
masses $m_{w}=0.067 m_e$ and $m_{b}=0.0919 m_e$\cite{ADA93}
in the Kronig-Penney model yielding a coupling $T_1=-0.0116$ meV.
We assume $\delta$-doping in the middle of the quantum wells.
The interaction with impurities located in different wells
is found to be negligibly small. In this section
screening is treated within the RPA assuming a free-electron gas.

\subsection{Density of states}
In Fig.~\ref{Figdichterpa} we show the resulting densities
of states for four different doping densities $N_D$.
For high $N_D$ the density of states
exhibits a monotonic increase from $\rho=0$ at $E\le E_{\rm min}$ to
$\rho\sim \rho_0$ for $E\to \infty$, where
$E_{\rm min}$ denotes the lowest edge of the density of states. 
In contrast to this the
density of states splits into two parts for small doping:
$\rho(E)$ takes finite values in a certain region
below $E=0$, which we will refer to as an impurity band.
For higher energies   $\rho(E)$ is quite similar to
the density of states of the free-electron gas.
These results are in good agreement 
with the findings of Ref.~\cite{SER89}. The onset  of the
impurity band occurs at slightly larger energies $E_{\rm min}$ here, 
as the wavefunctions are less confined due to 
the spreading into the barrier which was neglected in Ref.~\cite{SER89}.

We also marked the positions of the Fermi level $E^F$ (i.e., the 
chemical potential for $T_e=0$).
For low densities the position is just in the middle of
the impurity band, indicating that the impurity band
consists of exactly two states per impurity due to
the assumed spin degeneracy.  (This degeneracy would be lifted if
spin-resolved interaction was taken in account, 
see also appendix \ref{Apsigma}.)
For high densities the position of $E^F$ roughly equals
the Fermi level of the free-electron gas $N_D/\rho_0$.
The crossover between these two limits
occurs at $N_D\approx 5\cdot 10^{10}$/cm$^2$ where $E^F\approx 0$.

\begin{figure} 
\noindent
\epsfig{file=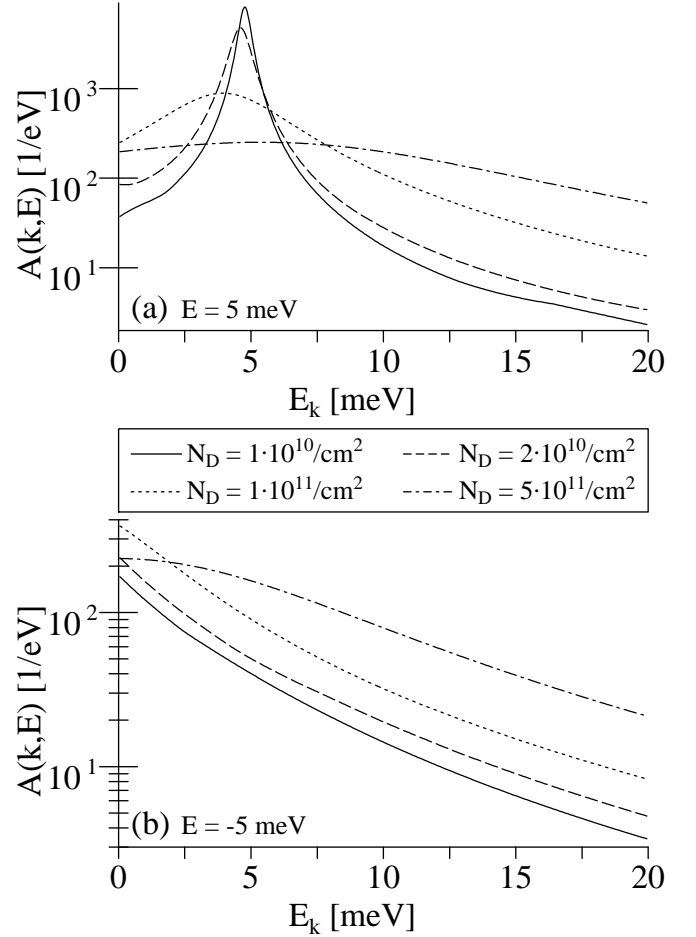,width=8.5cm}\\[0.2cm]
\caption[a]{Calculated spectral functions $A(E,k)$ versus $E_k=\hbar^2 k^2/2m$
using  RPA screening for different doping 
densities at $E=5$ meV (a) and $E=-5$ meV (b).}
\label{Figspektral}
\end{figure}

The respective spectral functions are plotted in Fig.~\ref{Figspektral}.
For $E=5$ meV $A(k,E)$ resembles a Lorentzian centered close
to $E_k\approx E$. This is the generic behavior of a free quasiparticle 
with a finite lifetime due to scattering. The width of the spectral
functions is increasing
with doping due to the enhanced scattering.
We find a full width at half maximum $\Gamma=0.5$ meV for 
$N_D\approx 1\cdot 10^{10}$/cm$^2$ and $\Gamma=5$ meV for 
$N_D\approx 1\cdot 10^{11}$/cm$^2$, which are in the range 
of the calculated values of $-2 {\rm Im}\{\Sigma({\bf k},E)\}$.

For $E=-5$ meV the spectral functions
exhibit a monotonic decrease. For high doping the slope
is comparable to the slope at $E=5$ meV. In contrast to this,
the spectral function for $E=5$ meV and $E=-5$ meV
are entirely different for low doping, indicating
that two different types of states occur. 
While the states for $N_D=10^{10}/$cm$^2$ 
are essentially free-particle states
at $E=5$ meV, they are localized in space for $E=-5$ meV,
which is the signature of an  impurity band\cite{GOL88}.

\subsection{Currents}

\begin{figure} 
\noindent
\epsfig{file=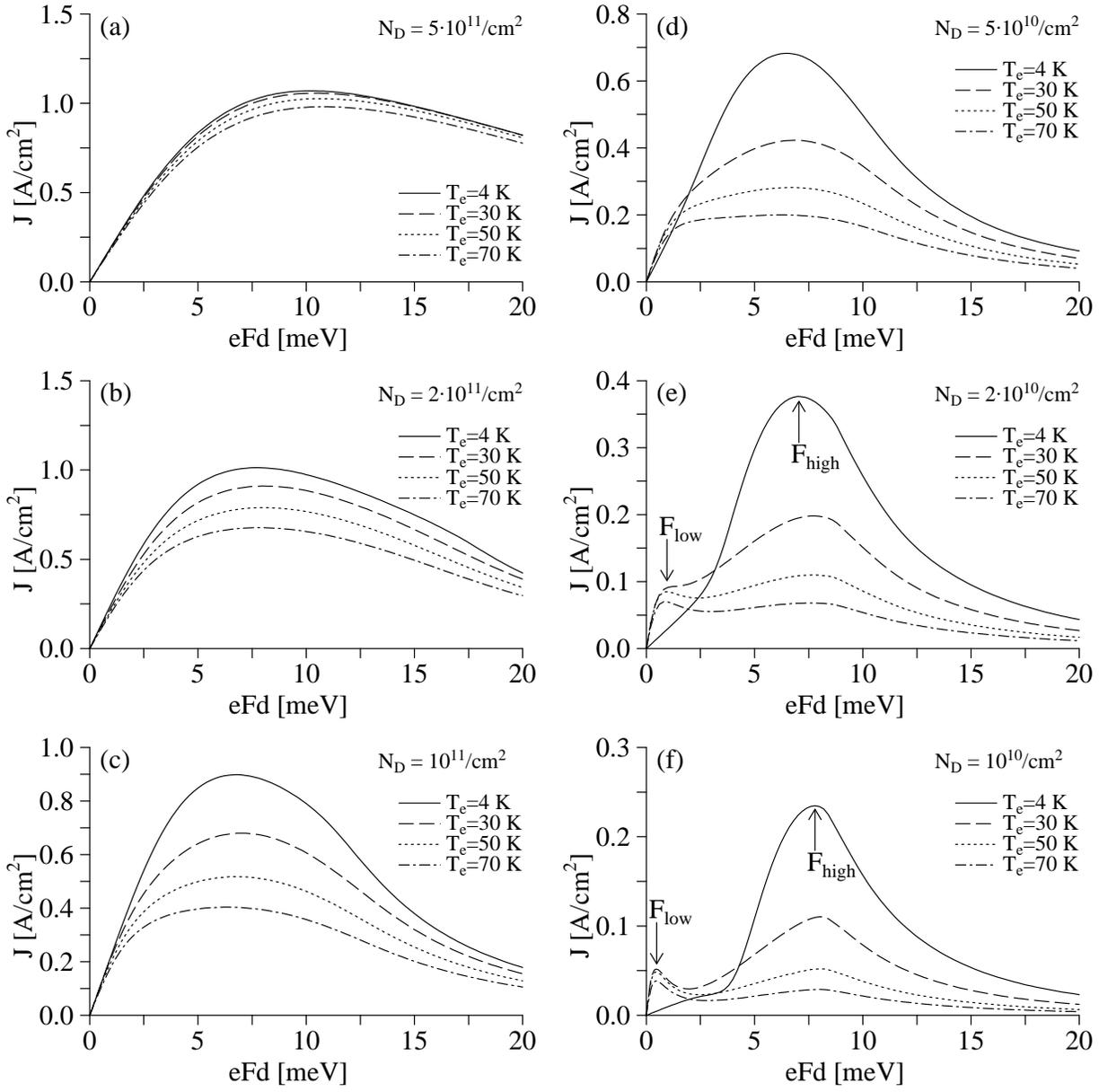,width=16cm}\\[0.2cm]
\caption[a]{Calculated temperature dependence of the
current-field relations for different doping densities.
The screening is treated within the RPA.}
\label{Figstromrpa}
\end{figure}

We calculate the current densities $J_{j\to j+1}(eFd)$ for
different electron temperatures $T_e$ from Eq.~(\ref{EqJ}).
The results are shown in Fig.~\ref{Figstromrpa}.
For all temperatures and densities we find an
ohmic range for low electric fields and negative differential conductance 
for high electric fields.
Let us first regard the high doping case (a-c), where no impurity
bands form and the Fermi level  is significantly above 
$E=0$.
In this case the approximation (\ref{EqJgamma}) is justified
and indeed we find a maximum at values of $eFd$ which are in the range
of calculated values of $\Gamma=-2 {\rm Im}\{\Sigma^{\rm ret}({\bf k},E)\} $.
The height is estimated by $J_{\rm max}= 0.91 $A/cm$^2$
from Eq.~(\ref{EqJmaxgamma}) which is in good agreement with the
full calculation at $T_e=4$ K. Note that the maximal current  is almost 
independent on the doping in this range.
For $N_D=5\cdot 10^{11}$/cm$^2$ the chemical potential
is larger than $k_BT_e$ for all temperatures. Thus, the current
is hardly affected by the temperature. In contrast to this
the current drops with temperature  
for lower doping ($N_D=10^{11}$/cm$^2$).
All these findings are in good agreement with the 
phenomenological description using a constant $\Gamma$ discussed
above.

\begin{figure} 
\noindent
\epsfig{file=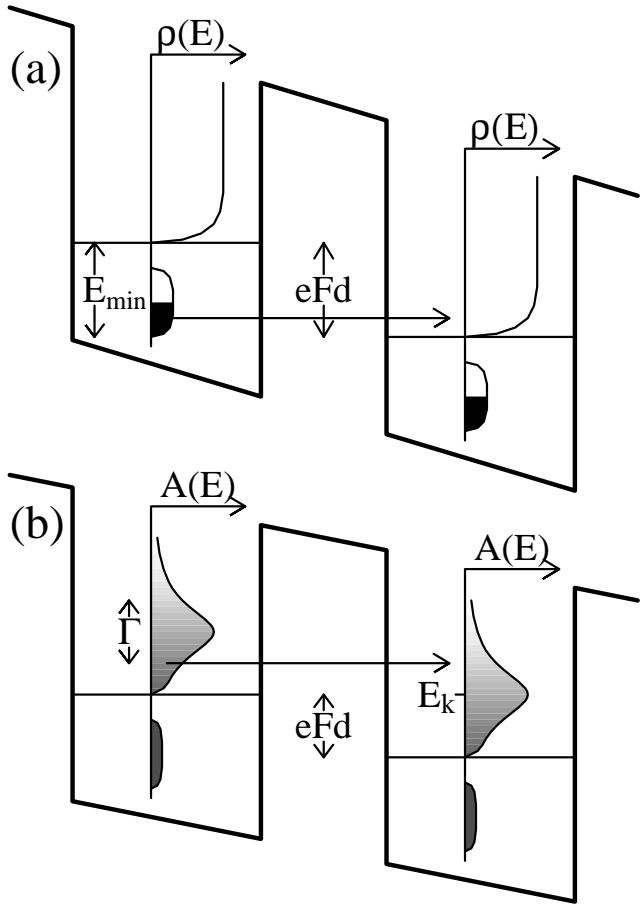,width=8.5cm}\\[0.2cm]
\caption[a]{Explanation of the two different
current maxima within a sketch of the conduction band profile:
(a) For low temperatures the electrons occupy the impurity band (black area).
As these states exhibit a flat spectral function (see Fig.~\ref{Figspektral}(b))
they contain contributions from essentially all
${\bf k}$-vectors and thus tunneling into the free-particle states
is possible at all energies. Maximal current is found  when all states
from the impurity band can tunnel into the free-particle states, i.e.,
$eFd\approx |E_{\rm min}|$.
(b) For high temperatures the electrons occupy the  free-electron states
as well (the grey scale indicates the occupation given by
the Fermi-function). The spectral function
$A({\bf k},E)$ of such a free-electron state with given wave 
vector ${\bf k}$ is peaked  around $E=E_k$ as shown in the figure. 
Due to ${\bf k}$-conservation
tunneling can only take place if the spectral functions 
for the same ${\bf k}$ of both wells overlap. On the other hand
a net current is caused by the difference in occupation.
This competition results in a current maximum for 
$eFd\approx \Gamma$ as shown in section III.}
\label{Figskizzeimp}
\end{figure}

For the low-doped samples (see Figs.~\ref{Figstromrpa}(e,f)) 
an entirely new scenario occurs. Here we find two different
maxima in the current-field relation whose relative weight
is changed by temperature.  The reason for this behavior is 
the presence of impurity bands for these doping levels.
For  $T_e = 4$ K we find a maximum  at $eF_{\rm high}d\approx 8$ meV.
This is due to tunneling from the impurity band to the free states 
(see Fig.~\ref{Figskizzeimp}(a)).
The maximum occurs at the energy where the bottom of the impurity band
in one well is aligned with the band edge of the free-electron states in the
neighboring  well, i.e., $eF_{\rm high}d\approx |E_{{\rm min}}|$.
An increasing temperature leads to a transfer of electrons
from the impurity band to the free-electron states and
consequently the current  at $eF_{\rm high}d$ decreases 
with increasing $T_e$. 
The density of states in the impurity band is much
lower than in the free-electron states, and hence
the majority of the electrons will be in the free-electron
states for  $k_BT_e\gtrsim |E_{{\rm min}}|$ (see Fig.~\ref{Figskizzeimp}(b)
where the grey scale denotes the relative occupation).
The current contribution due to the free-electron states 
can be understood within the phenomenological constant-$\Gamma$ approach.
There is a maximum at $eF_{\rm low}d\approx \Gamma$, which coincides with 
the full width at half maximum
of the spectral function at $E=5$ meV in Fig.~\ref{Figspektral}.
The amplitude of this maximum depends on two competing effects:
On the one hand the occupation of the free-electron states
increases with temperature. On the other hand the Fermi-factor
in Eq.~(\ref{EqJgamma}) strongly decreases with temperature.
This explains the calculated behavior, where the peak at $eF_{\rm low}d$
takes its maximum at intermediate temperatures.

\section{Results for Thomas-Fermi screening}
The properties related with the formation of impurity bands
are sensitive to the actual screening of the interaction\cite{GOL88}.
For low doping densities the density of states differs essentially
from the free-electron density of states and thus the use
of RPA-screening by a free-electron gas is questionable.
In order to take this effect into account we use 
the Thomas-Fermi approximation (TF) with the
actual density of states at the Fermi level 
(see appendix \ref{Apscreen}) in this section.
Of course neither the free-electron RPA nor the TF approximation
treat the screening entirely correctly, but we hope to obtain some
insight into the general features by comparing these two approaches.

\begin{figure} 
\noindent
\epsfig{file=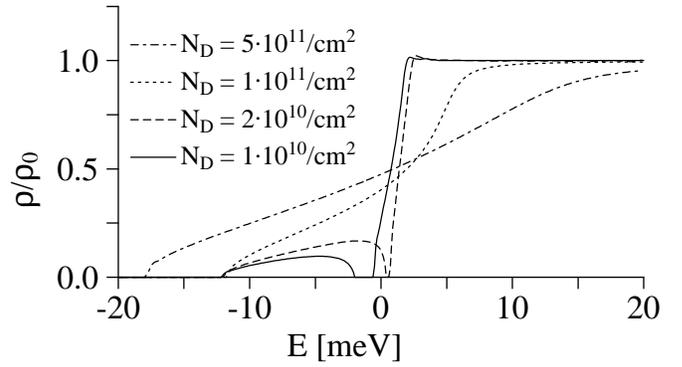,width=8.5cm}\\[0.2cm]
\caption[a]{Calculated density of states in units of the 2D 
free carrier density $\rho_0$ using  Thomas-Fermi screening 
for different doping densities.}
\label{Figdichtetf}
\end{figure}

In Fig.~\ref{Figdichtetf} we show the resulting density of states
which is in qualitative agreement with the results of the RPA screening
(Fig.~\ref{Figdichterpa}). For $N_D=5\cdot 10^{11}/$cm$^2$ the density
of states is almost identical while for  lower densities
some deviations occur. Especially the onset of the impurity band $E_{\rm min}$
is shifted to lower energies for TF-screening. Furthermore the impurity
bands extend over a larger energy range and have a lower density of states,
so that the total density is conserved. The reason for these deviations
lies in the fact that TF-screening is less effective than RPA screening 
if the actual density of states at the Fermi level is used. Therefore
both the binding energy of the impurities as well as the broadening
of the states become larger.

\begin{figure} 
\noindent
\epsfig{file=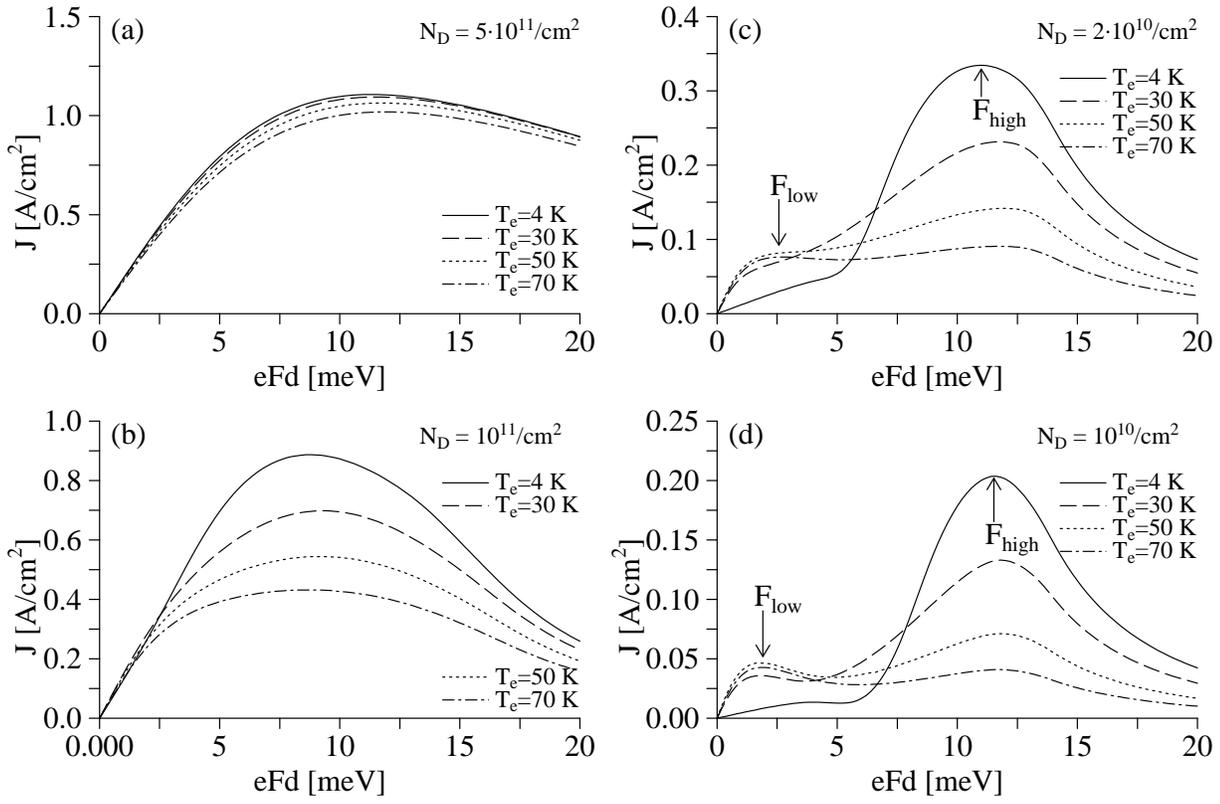,width=16cm}\\[0.2cm]
\caption[a]{Calculated temperature dependence of the
current-field relations for different doping densities using TF screening.}
\label{Figstromtf}
\end{figure}

This manifests itself in the calculated current densities 
(see Fig.~\ref{Figstromtf}). For high doping (a) the characteristics
are almost identical, while for lower doping deviations occur.
At first note that the maxima due to tunneling between free-electron states
(the maximum for $N_D=10^{11}/$cm$^2$ as well as the maxima $F_{\rm low}$ for
$N_D=2\cdot 10^{10}/$cm$^2$ and $N_D=10^{10}/$cm$^2$) are shifted to the
right according to the stronger scattering which increases $\Gamma$.
At second the peak at $F_{\rm high}$ is shifted to the right compared
to Fig.~\ref{Figstromrpa}.
Again we find $eF_{\rm high}d\approx E_{\rm min}$ for both densities.

Therefore we conclude that within both approximations for screening
the two maxima are determined by specific quantities describing the scattering.
$eF_{\rm low}d$ reflects the average broadening $\Gamma$ of the free-particle 
states and $eF_{\rm high}d$ is the energy separation between 
the onset of the impurity band and the free-particle states.

\section{Comparison with Experiments}
Previously\cite{WAC97b}, the formalism was applied 
for the highly-doped sample ($N_D=8.75\cdot 10^{11}/$cm$^2$)
of Refs.~\cite{HEL90,HEL91}.
Good quantitative agreement with the experimental
data was found, albeit using
a barrier width being 10\% smaller than the nominal value.
(A similar width had been used  in the original
analysis by the experimentalists as well\cite{HEL91}.)
The position of the first maximum occurred at $eFd=13$ meV, which
is almost independent of the barrier width (which mainly changes $T_1$)
and in excellent agreement with the experimental finding.
The second resonance, as well as the formation of field domains
was also studied in Ref.~\cite{WAC97b}, and again good agreement with 
experimental data was found.

A low-doped superlattice ($N_D=6\times 10^9/$cm$^2$,
$b=5$ nm, $w=15$ nm, $A=8\mu $m$^2$) with $N=10$ wells 
was used in the experiments of  Refs.~\cite{KEA95b,ZEU96}
in order to study the transport under strong THz irradiation
from a free-electron laser. Additional 
data for this sample will be given in the following. 
Without irradiation a broad maximum was found in the range
50 mV$<U<100$ mV where the current is almost constant. For $U>100$ mV
domain formation sets in. Dividing by the number of periods ($N=10$),
the maximum extends to $eF_{\rm unirr}d\approx 10$ meV.
In contrast to this, the photon-replica under strong THz-irradiation
could be consistently explained by assuming an \lq\lq instantaneous\rq\rq\ 
current-voltage characteristic \cite{ZEU96} with a distinct maximum at 
$U\approx 20$ mV (i.e., $eF_{\rm irr}d=2$ meV).

\begin{figure} 
\noindent
\epsfig{file=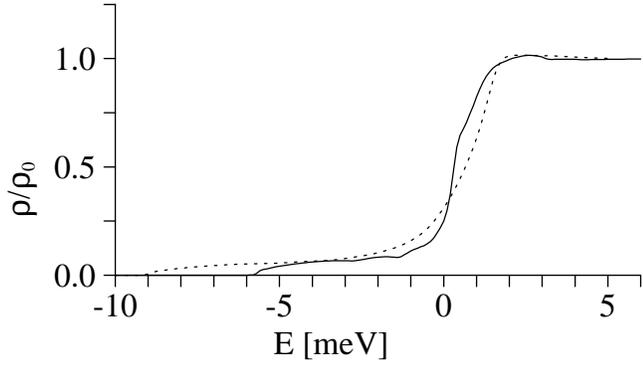,width=8.5cm}\\[0.2cm]
\caption[a]{Calculated density of states for the sample
parameters of Ref.~\cite{KEA95b,ZEU96} using RPA screening
(full line) and TF screening (dashed line).}
\label{FigdichteZ}
\end{figure}

We calculate the current-field relation for this superlattice
using the experimental sample parameters. 
In order to model the homogeneous doping we use  8 equally spaced 
$\delta$-doping layers per period. The calculated density of states
for both RPA and TF screening is shown in Fig.~\ref{FigdichteZ}.
The density of states resembles the result
for low doping found before. Nevertheless we do not find a separation
between the impurity band and the free-particle states.
The reason is the homogeneous doping: The different impurity
positions have different binding energies which smears the impurity band.
Again the onset of the impurity band occurs at significantly lower energies
within the reduced  Thomas-Fermi screening. Both values
of $|E_{\rm min}|$ are smaller than the corresponding values for low doping
for the calculation done before 
(see Figs.~\ref{Figdichterpa},\ref{Figdichtetf}).
This is due to the larger well width in the sample: 
The Wannier-states
are less localized and therefore the matrix-element for impurity 
scattering(\ref{Eqimpmat}) as well as the binding energy of the impurity levels
is smaller.

\begin{figure} 
\noindent
\epsfig{file=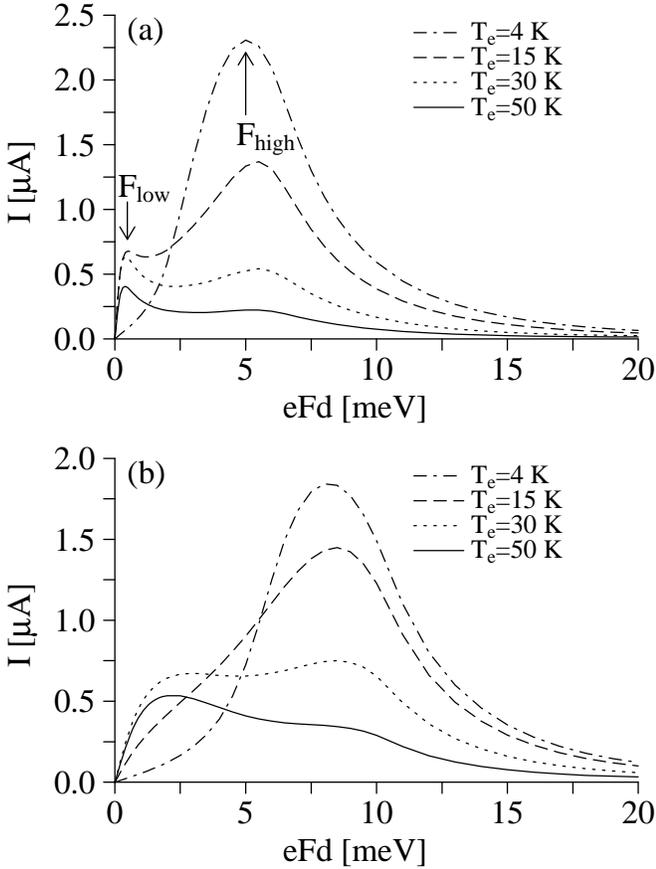,width=8.5cm}\\[0.2cm]
\caption[a]{Calculated temperature dependence of the
current-field relations for the sample of Ref.~\cite{KEA95b,ZEU96}.
The screening is treated within the RPA (a) and within the
Thomas-Fermi approximation (b).}
\label{FigstromZ}
\end{figure}

The results for the current-field relation are shown in Fig.~\ref{FigstromZ}.
Again we find two maxima whose relative height changes with temperature.
The position of the maximum for low temperatures, 
$eF_{\rm high}d$, is almost identical to the value
of $|E_{\rm min}|$ for both types of screening like in the calculations 
shown before.

Now we can offer an explanation for the two different
maxima occurring in the experiment \cite{KEA95b,ZEU96} 
with and without irradiation mentioned above.
For low electron temperatures and
without irradiation the maximum at $eF_{\rm high}d$
dominates the transport and thus domain formation  sets in at voltages 
exceeding $U\approx NeF_{\rm high}d$ where $N=10$ is the number of wells. 
If the THz radiation is present
the electrons are excited from the impurity band into the free-electron states
corresponding to a larger effective electron temperature.
Thus, the maximum at $U=NeF_{\rm low}d$ 
is dominant, and the photon replicas corresponding to this feature
are seen experimentally.
The experimental values therefore suggest $eF_{\rm high}d=10$ meV and
$eF_{\rm low}d=2$ meV which is in excellent agreement with 
the calculation using Thomas-Fermi screening.
This indicates that that the RPA using a free electron gas overestimates
the screening in low doped samples. The Thomas Fermi approximation
with the actual density of states at the Fermi level seems
to reproduce the experimental results better
in agreement with our argumentation in appendix \ref{Apscreen}.
Therefore we will use it in the following for comparison 
with the experiment.

\begin{figure} 
\noindent
\epsfig{file=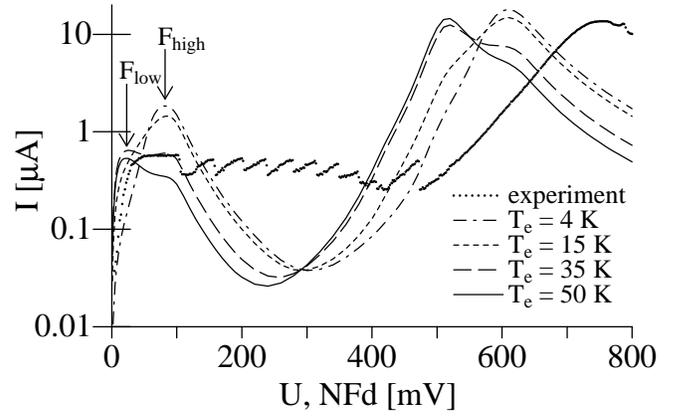,width=8.5cm}\\[0.2cm]
\caption[a]{Experimental current-voltage characteristic 
without external irradiation
together with calculations for  
different electron temperatures.
In the calculation we estimate the
bias by $NFd$ assuming a homogeneous field distribution and
neglecting potential drops in the contact regions.}
\label{Figtwomax}
\end{figure}

In Fig.~\ref{Figtwomax} we compare  the  calculated 
currents with the 
experimental current-voltage characteristic without irradiation in a wider
range of fields. Here we included the resonance around  $eFd\approx 50$ meV
between the lowest level and the first excited level as well.
The calculation of the corresponding current is completely 
analogously to  Eq.~(\ref{EqJ}), for details see Ref.~\cite{WAC97}.
Note that there are no fitting parameters involved in the calculation --
all quantities including matrix elements and spectral functions
are directly calculated for the given sample parameters as outlined above.
Let us first focus on the low field region.
For $U<10$ mV the experimental data are in good agreement
with the calculated currents for $T_e=4$ K, the experimental 
lattice temperature. With increasing bias, the experimental data
leave the 4K curve and follow the $T_e=35$ K curve at the plateau
between 50 mV and 100 mV. This can be understood by electron heating
inside the sample:
For a voltage drop of 8 mV per period and a current of $0.6\mu$A
the Joule heating is $P\approx 10$ pW per electron.
In a recent transport experiment a distribution function with
$T_e\approx 40$ K was observed\cite{HIL96} for this amount of heating
albeit using a sample with higher doping.
This shows that the electron temperature
strongly deviates from the lattice temperature in the experiment
considered in good agreement with our findings.
At $U>100$ mV, the homogeneous field distribution becomes unstable,
as the region of negative differential conductance is reached
and electric field domains form, causing the saw-tooth shape of the 
characteristic
(see Ref.~\cite{WAC97} and references cited therein).
For $U>450$ mV one can clearly see the resonance between
the lower level and the first excited level in the well which is located
48 meV above the ground level.
Again the calculation exhibits two different peaks depending on the electron 
temperature due to the different occupation of the impurity bands, although
only the high temperature result should be meaningful due to
the heating of the carriers.
The peak height of around $14\mu$A is in
excellent agreement with the 
value of $13.6\mu$A found experimentally for our sample.
The experimental peak position is located at a higher bias.
This may be due to a voltage drop in the receiving
contact, where a low-doped spacer layer of 
$d_{\rm contact}=50$ nm thickness exists.
If the electric field inside the sample is large, it cannot
be screened within the spacer layer and the effective field
inside the sample is $U\sim (Nd+d_{\rm contact})F$ instead of
$U=NFd$ as assumed in the figure.

\begin{figure} 
\noindent
\epsfig{file=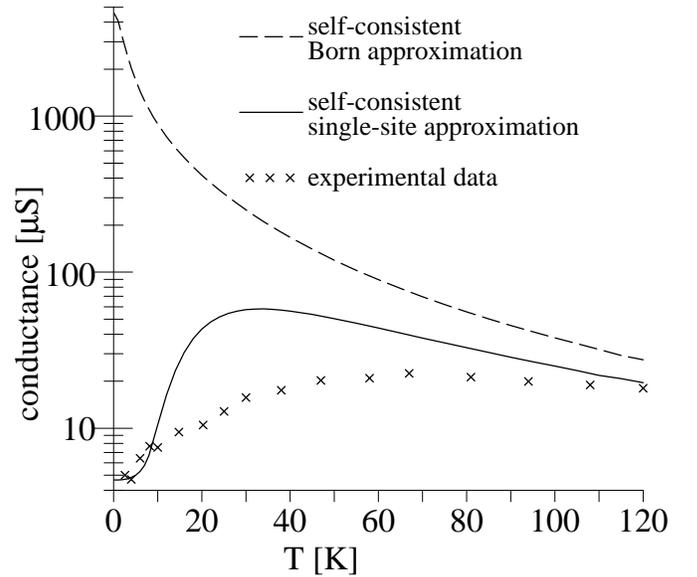,width=8.5cm}\\[0.2cm]
\caption[a]{Temperature dependence of the
zero-bias conductance for the sample of Refs.~\cite{KEA95b,ZEU96}.
Full line: Calculation  using spectral functions from the
single-site approximation and TF screening, 
dashed line: Calculation  using spectral 
functions from the self-consistent Born-approximation,
crosses: experimental data}
\label{Figmobil}
\end{figure}

In order to circumvent the problems of electron heating
we have investigated the temperature dependence of the zero-bias conductance
$G=dI/dU$, where $T_e$ should be equal to the lattice temperature $T$. 
The results are shown in Fig.~\ref{Figmobil}
both for our full calculation using TF-screening as well as for spectral
functions calculated within the 
self-consistent Born-approximation (\ref{EqBorn}) 
where no impurity bands form.
In the latter case $G$ is monotonously decreasing
in $T$ as shown in Fig.~\ref{Figmobil}. This can be easily
understood within the phenomenological constant-$\Gamma$ 
approach(\ref{Eqcond}).
However, a different scenario emerges if
the electrons occupy impurity bands for
low temperatures. Then  $G$ is strongly suppressed due to 
the small values of $A({\bf k},E)$ for $E<0$, see Fig.~\ref{Figspektral}.
As temperature is increased, more
electrons are excited to the free-electron states, and $G$ {\em increases}
with $T$ until the impurity bands are almost
empty at  $k_BT\sim |E_{{\rm min}}|$.  
This physical picture is 
confirmed by the experimental data shown in  Fig.~\ref{Figmobil}.
At low temperatures the agreement is quantitative,
while at intermediate $T$ the theory overestimates $G$; 
this is most likely due to additional scattering processes not 
included in our calculation, or by the presence of a contact 
resistance which may limit the experimental conductance.

Thus we may conclude that the results of our calculations are in good
agreement with experimental data both for high and low doping.
Nevertheless, a direct observation of the two-peak structure is not 
available so far.

\section{Photon-assisted tunneling}
The standard theory of PAT considers tunneling between two reservoirs
between which a dc-voltage $\Delta U_{\rm dc}$ is applied.
Let us denote the current-voltage 
characteristic (IV) without irradiation by
$I^{\rm dc}(\Delta U_{\rm dc})$. Under irradiation  an additional ac-bias 
$U=\Delta U_{\rm ac}\cos(2\pi \nu t)$ is induced between the reservoirs.
Then the time-averaged IV is given by\cite{TUC85}:
\begin{equation}
I^{\rm irr}(\Delta U_{\rm dc})=\sum_{l=-\infty}^{\infty}[J_l(\alpha)]^2
I^{\rm dc}\left(\Delta U_{\rm dc}+\frac{lh\nu}{e}\right)\;,
\label{EqTucker}
\end{equation}
where $\alpha=e\Delta U_{\rm ac}/(h\nu)$ and $J_l$ is the ordinary Bessel
function of order $l$. Thus, the current under irradiation is given as
a sum over the photon replicas $e\Delta U_{\rm dc}+lh\nu$ where the 
alignment of the wells is shifted by integer multiples of the
photon energy. The great advantage of Eq.~(\ref{EqTucker}) is that it
expresses all transport properties in terms of $I^{\rm dc}(\Delta U_{\rm dc})$.
Eq.~(\ref{EqTucker})  has been
applied to photon-assisted tunneling in weakly-coupled superlattices
by identifying $\Delta U_{\rm dc}$=$F_{\rm dc}d$ and
$\Delta U_{\rm ac}=F_{\rm ac}d$, where $F_{\rm ac}$ is the 
field component of the microwave field in the growth direction of the 
superlattice\cite{KEA95a,INA96,ZEU96}.
Note that Eq.~(\ref{EqTucker}) also holds within a 
miniband model for {\em strongly-coupled} superlattices\cite{IGN95}.

\begin{figure}
\noindent\epsfig{file=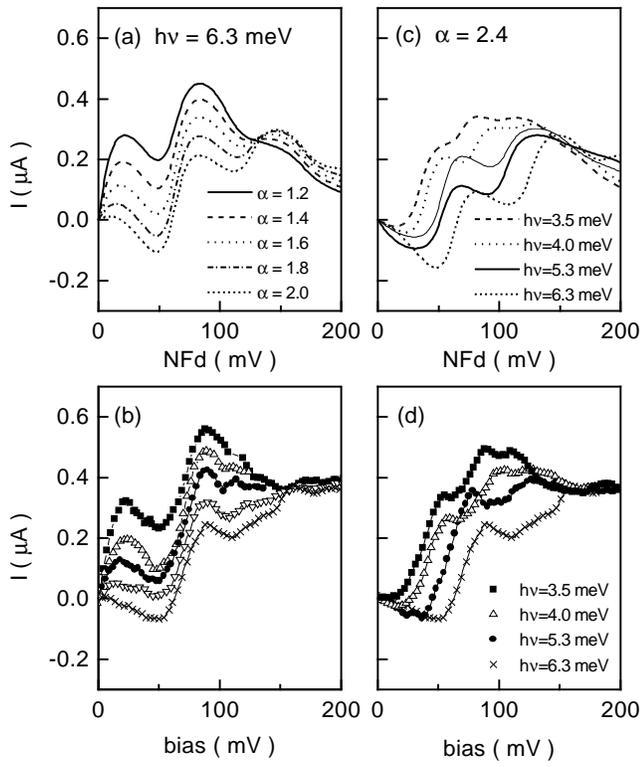,width=8.5cm}\\[0.2cm]
\caption[a]{Current-voltage characteristics  under irradiation.
a) Theoretical results for $h\nu =6.3$ meV and different
field strength $eF_{ac}d=\alpha h\nu $ of the irradiation.
b) Experimental results for  $ h\nu=6.3$ meV and different
laser intensities increasing from the top to the bottom.
The actual  values $F_{ac}$ inside the sample are not accessible.
c) Theoretical results  for $\alpha=2.4$ and different
photon energies. The thin line depicts $h\nu=5.3$ meV and 
$\alpha=2.1$. d) Experimental results for different photon energies.
The laser intensity  was tuned to give maximum  negative conductance.}
\label{Figeinstrahl}
\end{figure}

In Refs.~\cite{KEA95b,PLA97} the $I^{\rm dc}\left(eF_{\rm dc}d\right)$ 
curve was  calculated  phenomenologically 
and a  qualitative agreement with the experimental data could be obtained.
To refine the theory we use the results
of our  microscopic calculation (see Fig.~\ref{Figtwomax}) here.
As the external irradiation heats the electronic distribution
for zero bias as well, we use the curve for $T_e=35$ K. 
At this electron temperature about 50\% of the electrons are 
occupying the states in the impurity band. Of course the actual
electron distribution may deviate from a Fermi distribution
under the strong irradiation. Nevertheless we expect that
an effective temperature approach gives a reasonable description
of the excited carriers. Further calculations show that
the results for $T_e=30$ K or $T_e=40$ K do not differ qualitatively.
Quantitative agreement between theory (Fig.~\ref{Figeinstrahl}a) 
and experiment (Fig.~\ref{Figeinstrahl}b) is found for
$h \nu=6.3$ meV (1.5 THz) for different strengths of the laser field.
We find a direct tunneling  peak at 
$U_{\rm dir}=N F_{\rm low}d\approx 20$ mV and photon replicas 
at $U\approx U_{\rm dir}+Nh\nu/e=83$ mV  and 
$U\approx U_{\rm dir}+2Nh \nu/e=146$ mV. 
Photon-replicas of the second peak around 
$NF_{\rm high}d\approx 100$ mV are less pronounced as this peak
is broader. Our calculations show that they become visible 
if larger photon energies are used.
For low bias and high 
intensities there is a region of absolute negative 
conductance\cite{KEA95b}, which we focus on in the following.

In Fig.~\ref{Figeinstrahl}(d) the laser intensity has been tuned
such that maximal absolute negative conductance occurred for 
each of the  different laser frequencies.
Then we observe a minimum in the current at 
$U\approx -U_{\rm dir}+ Nh\nu /e$ which is just the first 
photon replica of the direct tunneling peak on the negative bias side. 
This replica dominates the current if the direct tunneling channel is 
suppressed close to the zero of $J_0(\alpha)$ in Eq.~(\ref{EqTucker}),
i.e., $\alpha\approx 2.4$, as used in the calculation 
(Fig.~\ref{Figeinstrahl}c). 
Both the theoretical and experimental results show that
absolute negative conductance persists in a wide 
range of frequencies but becomes less pronounced 
with decreasing photon energy.
In the calculation absolute negative conductance vanishes for
$h\nu <1.8$ meV which is approximately equal to $h\nu \lesssim eF_{\rm low}d$.
The latter relation has been verified by calculations 
for different samples as well.
For $h \nu=5.3$ meV  a smaller value 
of $\alpha=2.1$ (thin line) agrees better with the experimental data
(in the same sense the value $\alpha=2.0$ agrees better for 
$h \nu=6.3$ meV, compare Fig.~\ref{Figeinstrahl}b).
This may be explained as follows:
If strong NDC is present in doped superlattices the homogeneous 
field distribution becomes unstable and either self-sustained
oscillations or stable field domains form \cite{WAC97a}.
Then the IV  deviates from the 
relation for homogeneous field distribution, where $U=NFd$, and typically
shows less pronounced NDC. Therefore maximal negative conductance 
is observed at a laser field corresponding to a value of $\alpha<2.4$,
where the NDC is weaker and the  field distribution is still homogeneous. 
The presence of an  inhomogeneous field distribution  could explain the
deviations between theory and experiment for $U>150$ mV as well.

\section{Discussion}
We have investigated the electrical transport
in weakly-coupled  doped superlattices, where the transport
is given by sequential tunneling.
Our calculations give negative differential conductance for all 
doping densities and temperatures at sufficiently large electric fields.
This will give rise to instabilities leading to domain formation 
\cite{PRE94,BON94}
or self-sustained current oscillations \cite{KAS97}. Within the
full transport model using Eq.~(\ref{EqJ}) 
these effects are discussed in  Ref.~\cite{WAC97}.

For high doping $N_D\gtrsim 10^{11}$/cm$^2$ 
or high temperatures $k_BT\gtrsim |E_{\rm min}|$ the electrons 
mainly occupy free quasiparticle states. Then the general
behavior can be understood within a phenomenological model
using a constant self-energy $-i\Gamma/2$.
The current exhibits a maximum at $eFd\approx \Gamma$ which can be
used to investigate scattering processes.
For doped samples impurity scattering is an important scattering process
which we considered here.
The inclusion of further scattering processes like 
interface roughness scattering, electron-electron scattering,
or phonon scattering  will increase $\Gamma$ and therefore
the position of the first peak. 

For low-doped samples $N_D\ll  10^{11}$/cm$^2$ and low temperatures   
$k_BT\ll |E_{\rm min}|$ the presence of impurity bands 
influences significantly the low-field transport.
Then a second maximum at $eF_{\rm high}d\approx |E_{\rm min}|$
occurs. This maximum  provides a possibility to obtain information about
the position of the impurity band. This position depends strongly
on the screening as can be seen by comparison of the calculations
within RPA and TF. Therefore such experiments could serve 
as a test on various models for screening. 
For the sample considered here $E_{\rm min}$ calculated
within the free-particle RPA is too low compared with the
experimental onset of domain formation. In contrast the
result using TF screening gives excellent agreement with the
experimental data. This indicates that screening within the 
free-particle RPA is too strong for low-doped samples.

Furthermore it would be interesting to see if effects due to
spin splitting of the impurity band are visible in experiments.
These experiments can be both carried out in doped superlattices as well
as in resonant tunneling between neighboring two-dimensional electron gases
in the spirit of Ref.~\cite{MUR95,TUR96}. The latter has the advantage that 
problems due to domain formation in the region of negative 
differential conductivity are absent.
An important aspect in such experiments is the problem of electron heating
as the temperature $T_e$ refers to the temperature of the electronic distribution.
In order to avoid heating, 
structures with thick barriers should be 
used where the Joule heating becomes small.

For external irradiation we have demonstrated both experimentally
and theoretically that absolute negative differential conductance 
persists in a wide range of frequencies $h\nu \gtrsim eF_{\rm low}d$.
The calculated current-voltage characteristics are in excellent 
quantitative agreement with the experimental data using
a microscopically calculated $I^{\rm dc}(eFd)$ combined with the
Tucker formula (\ref{EqTucker}). 
Recently, the same model has been applied to
photon-assisted tunneling in a different sample\cite{WACp}
and  quantitative agreement has been achieved as well.
This shows that
the simple Tucker formula allows for a quantitative 
description of photon-assisted transport in weakly-coupled
superlattices.

\acknowledgements
We want to thank Ben Hu for stimulating discussions.
A.W. and S.Z. acknowledge financial support by the
Deutsche Forschungsgemeinschaft.
Work performed at the Center for Terahertz Science and Technology was
supported by the Office of Naval Research, the Army Research Office
and the National Science Foundation.

\appendix
\section{Calculation of the transition elements}\label{ApT1}
In a superlattice structure the 
coupling between neighboring  wells $T_1$ is related to the
dispersion relation $E(q)$ of the miniband (see \cite{WAC97}) via
\begin{equation}
T_1=\frac{d}{2\pi}\int_{-\pi/d}^{\pi/d}dq E(q)e^{iqd}\, .
\end{equation}
For a next-neighbor tight-binding model we have
$E(q)=-(\Delta/2)\cos(qd)$ and $T_1$ is equal to 
a fourth of the miniband width $\Delta$.
Here we calculate $E(q)$ for a given superlattice via the 
Kronig-Penney model for the respective sample parameters.
Furthermore we determine the Wannier-functions $\Psi(z-jd)$ localized
in well $j$ from the Bloch-functions $\phi_q(z)$, where we choose the
phase of the Bloch functions such that the Wannier functions are
maximally localized\cite{KOH59}.
These Wannier-functions are used for the calculation of the
matrix elements for scattering.

\section{Calculation of the self-energies}\label{Apsigma}
\begin{figure} 
\epsfig{file=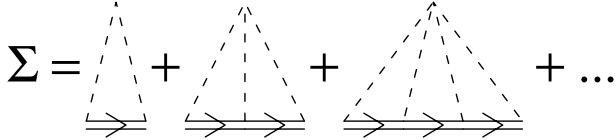,width=8.5cm}\\[0.2cm]
\caption[a]{The
self-consistent single-site approximation. The dashed lines
indicate impurity potentials and the double lines denote the
full Green-function}
\label{Figssa}
\end{figure}

We assume that the electron density in the conduction band is provided 
by doping of the superlattice. Thus, scattering at ionized impurities
is an important scattering process. In addition there may be interface 
roughness scattering, phonon scattering, or electron-electron 
scattering, which we will neglect in the following.
For weakly-coupled superlattices, the dominating
scattering process occurs within the wells, which are assumed to be identical.
Thus, the well index $j$ can be omitted. Scattering at 
the ionized impurities is described by the Hamiltonian
\begin{equation}
\hat{H}^{\rm scatt}=\frac{1}{A}\sum_{{\bf k},{\bf p},\alpha}
V_{\alpha}({\bf p})a^{\dag}({\bf k}+{\bf p})a({\bf k})\label{EqH0imp}
\end{equation}
where the subscript $\alpha$ denotes the impurity
located at the position  $({\bf r}_{\alpha},z_{\alpha})$. 
The matrix element is calculated with the Wannier functions yielding:
\begin{eqnarray}
V_{\alpha}({\bf p})&=&\int d^2r\,dz\,e^{-i{\bf p}\cdot{\bf r}}
\Psi^*(z)\Psi(z)\nonumber \\
&&\quad \times \frac{-e^2}{4\pi\epsilon_s\epsilon_0
\sqrt{|{\bf r}-{\bf r}_{\alpha}|^2+(z-z_{\alpha})^2}}\nonumber \\
&=&\frac{-e^2}{2\epsilon_s\epsilon_0 p}\int dz\,\Psi^*(z)\Psi(z)
e^{-p|z-z_{\alpha}|}
e^{-i{\bf p}\cdot{\bf r_{\alpha}}}\, .\label{Eqimpmat}
\end{eqnarray} 
Using the bare Coulomb interaction $V_{\alpha}({\bf p})$ the
relevant integrals in the self-energies
are divergent. Thus screening
is essential for the calculation. We treat screening within
the Random-Phase approximation (RPA) of the free-electron gas as well
as within an effective Thomas Fermi approximation(TF) 
(see appendix \ref{Apscreen}). 
With the screened impurity potential 
$V^{\rm sc}_{\alpha}({\bf p})=V_{\alpha}({\bf p})/\epsilon({\bf p})$
the self-energy is calculated within the
self-consistent single-site
approximation (shown diagrammatically in 
Fig.~\ref{Figssa}) like in Ref.~\cite{GOL88}.
Then the self-energy contribution from the impurity $\alpha$ is given by
\begin{eqnarray}
\Sigma_{\alpha}({\bf k},E)&=&\frac{1}{A^2}\sum_{{\bf k}_1}
V_{\alpha}^{{\rm sc}}({\bf k}-{\bf k}_1) G({\bf k}_1,E)
V_{\alpha}^{{\rm sc}}({\bf k}_1-{\bf k}) \nonumber \\
&+&\frac{1}{A^3}\sum_{{\bf k}_1,{\bf k}_2}
V_{\alpha}^{{\rm sc}}({\bf k}-{\bf k}_1) G({\bf k}_1,E)
V_{\alpha}^{{\rm sc}}({\bf k}_1-{\bf k}_2)\nonumber \\
&&\phantom{\frac{1}{A^3}\sum_{{\bf k}_1,{\bf k}_2}} \times G({\bf k}_2,E)
V_{\alpha}^{{\rm sc}}({\bf k_2}-{\bf k})\nonumber \\
&+&\dots \label{Eqsigmaalpha1}
\end{eqnarray}
where $G({\bf k},E)=(E-E_k-\Sigma^{{\rm ret}}({\bf k},E))^{-1}$
is the full retarded Green function and
\begin{equation}
\Sigma^{{\rm ret}}({\bf k},E)=\sum_{\alpha}\Sigma_{\alpha}({\bf k},E)
\label{Eqsigmasum}
\end{equation}
is the sum over all contributions.
In case of $\delta$-doping the impurity potentials (\ref{Eqimpmat})
from different impurities located in  the same well  differ
only by a phase factor and the sum over $\alpha$ can be replaced by
a multiplication with the number of impurities per layer $N_DA$.
Eq.~(\ref{Eqsigmaalpha1}) can be transformed to the self-consistent equation 
(see, e.g., \cite{GOL88})
\begin{eqnarray}
K_{\alpha}({\bf k}_1,{\bf k},E)&=&V_{\alpha}^{{\rm sc}}({\bf k}_1-{\bf k})
\label{EqK} \\
&+&\frac{1}{A}\sum_{{\bf k}_2}
V_{\alpha}^{{\rm sc}}({\bf k}_1-{\bf k}_2) G({\bf k}_2,E)
K_{\alpha}({\bf k}_2,{\bf k},E)\nonumber
\end{eqnarray}
which we solve numerically
for a given self-energy function $\Sigma^{{\rm ret}}({\bf k},E)$ 
entering $G({\bf k}_2,E)$. We parametrize ${\bf k}_1,{\bf k}$ by
$E_{k_1},E_k$, and $\phi=\angle({\bf k}_1,{\bf k})$ and discretize the
resulting equation. This gives a set of 
linear equations for the components
of $K(E_{k_1},\phi)$ which is solved by matrix inversion.
Then the self-energy reads:
\begin{equation}
\Sigma_{\alpha}({\bf k},E)=
\frac{1}{A^2}\sum_{{\bf k}_1}
V_{\alpha}^{{\rm sc}}({\bf k}_1-{\bf k}) G({\bf k}_1,E)
K_{\alpha}({\bf k}_1,{\bf k},E)\, .
\label{Eqsigmaalpha}
\end{equation}
The equations (\ref{Eqsigmasum},\ref{EqK},\ref{Eqsigmaalpha}) 
have to be solved
self-consistently thus determining the self-energy 
$\Sigma^{{\rm ret}}({\bf k},E)$.

Our single-site-approximations neglects all contributions from
crossed diagrams (leading to weak-localization effects, as 
considered in Ref.~\cite{SZO92}) as well as the 
spin-resolved electron-electron
interaction  leading to the splitting of the
impurity bands (the Mott transition, see e.g. \cite{SHK84}).
The latter may become important for very low densities 
when the impurity bands are narrow.
Within this approximation for $\Sigma^{{\rm ret}}({\bf k},E)$
the integral (\ref{Eqdichte}) is a well defined quantity, as
${\rm Im}\{\Sigma^{{\rm ret}}({\bf k},E)\}= 0$ 
(and thus $A({\bf k},E)=0$)
for $E<E_{\rm min}$.

Finally note that no impurity bands are found within the
self-consistent Born approximation
\begin{equation}
\Sigma_{\alpha}({\bf k},E)=\frac{1}{A^2}\sum_{{\bf k}_1}
V_{\alpha}^{{\rm sc}}({\bf k}-{\bf k}_1) G({\bf k}_1,E)
V_{\alpha}^{{\rm sc}}({\bf k}_1-{\bf k}) \label{EqBorn}
\end{equation}
which is just the first diagram from Fig.~\ref{Figssa}.

\section{The screening}\label{Apscreen}
In order to consider screening we have to include the electron-electron
interaction given by the Hamiltonian
\begin{eqnarray}
\hat{H}^{ee}&=&\frac{1}{2A}\sum_{{\bf k},{\bf k'},{\bf p}}
W({\bf p})a^{\dag }({\bf k}+{\bf p})a^{\dag}({\bf k'}-{\bf p})
a({\bf k'})a({\bf k})\, .
\end{eqnarray}
where the matrix element is calculated via
\begin{eqnarray}
W({\bf p})=
\frac{e^2}{2\epsilon_s\epsilon_0 p}\int dz_1\int dz_2\,
\Psi^{*}(z_1)\Psi^{*}(z_2)\nonumber\\
\times \Psi(z_2)\Psi(z_1)e^{-p|z_1-z_2|}\, .
\end{eqnarray}
Within the random-phase approximation (RPA)
the screening of the impurity potentials is described by\cite{MAH90}
\begin{eqnarray}
V_{\alpha}^{\rm RPA}({\bf p})=
\frac{V_{\alpha}({\bf p})}{1-\Pi^0({\bf p},\omega=0)W({\bf p})}\, .
\label{EqVscreened}
\end{eqnarray}
For a free-electron gas  the two-dimensional  vacuum
polarizability $\Pi^0({\bf p},\omega=0)$ for $T=0$
is given by \cite{STE67a}
\begin{equation}
\Pi^{0}({\bf p},\omega=0)=-\rho_0 \left(1-\Theta(p-2k_F)
\sqrt{1-4\frac{k_F^2}{p^2}}\right)\, .\label{Eqpolarization}
\end{equation}
where $k_F=(2\pi N_D)^{1/2}$ is the Fermi wave-vector.

\begin{figure} 
\noindent
\epsfig{file=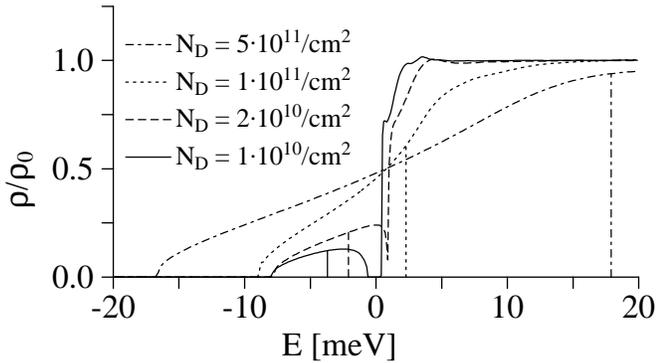,width=8.5cm}\\[0.2cm]
\caption[a]{Calculated density of states in units of the 2D 
free carrier density $\rho_0$ using RPA screening. 
The vertical lines indicate the position of the chemical
potential for $T=0$ at the respective doping densities.}
\label{Figdichterpa}
\end{figure}

Actually, the electronic states are
affected by the impurity scattering, which may change the density of states
dramatically as can be seen from Fig.~\ref{Figdichterpa}.
Now the polarizability  $\Pi(p=0)$ is related to the 
{\em actual density of states} at 
the chemical potential  which is
significantly lower than $\rho_0$.
Calculations within the Born-approximation
show that the $p$-dependence of the polarizability becomes weaker
and that $\Pi(0)$ decreases with increasing scattering \cite{AND82a,DAS83}.
In order to accommodate these trends we make the replacement
$\Pi^0(k)\to\Pi^*(k) = -\rho(E^F)$, given by the calculated
density of states of Fig.~\ref{Figdichterpa} and the
chemical potential at $T=0$. Then we obtain the screened impurity interaction
\begin{eqnarray}
V_{\alpha}^{\rm TF}({\bf p})=
\frac{V_{\alpha}({\bf p})}{1+\rho(E^F)W({\bf p})}\, .
\end{eqnarray}
This is equivalent to the
Thomas-Fermi approximation for the screening.
The same type of screening has been considered in Ref.~\cite{GOL88} as well.
Of course both ways of including screening are approximations.
In a full calculation the scattering has to be treated
self-consistently in the calculation of the polarizability.
Such a calculation was performed in Ref.~\cite{HU93} for a 
quantum wire within the restriction of a delta-potential
for impurity scattering.

Eq.~(\ref{EqVscreened}) only considers  screening  within the same
well. The extension to screening by electrons from
neighboring wells is given  in section 6 of Ref.~\cite{WAC97}.
The results are almost indistinguishable for the samples with thick barrier
width considered in sections IV and V (see also Ref.~\cite{WAC97b} for
the screened matrix elements).  Screening by electrons from
neighboring wells becomes more important for a smaller barrier width
as used in section VI and VII where the formalism from Ref.~\cite{WAC97}
was used with the polarization (\ref{Eqpolarization}) for RPA and
$\Pi^0(k)= -\rho(E^F)$ for the TF case.
The temperature dependence of the screening is neglected in all calculations.


\begin{thebibliography}{10}
\bibitem[*]{byline}  Present address: Klarastr. 5a, 80636 M{\"u}nchen, Germany.

\bibitem{ESA74}
L. Esaki and L.~L. Chang, Phys.~Rev.~Lett. {\bf 33},  495  (1974).

\bibitem{CAP86}
F. Capasso, K. Mohammed, and A.~Y. Cho, Appl.~Phys.~Lett. {\bf 48},  478
  (1986).

\bibitem{KWO95}
S.~H. Kwok, H.~T. Grahn, M. Ramsteiner, K. Ploog, F. Prengel, A. Wacker, E.
  Sch{\"o}ll, S. Murugkar, and R. Merlin, Phys.~Rev.~B {\bf 51},  9943  (1995).

\bibitem{KAS97}
J. Kastrup, R. Hey, K.~H. Ploog, H.~T. Grahn, L.~L. Bonilla, M. Kindelan, M.
  Moscoso, A. Wacker, and J. Gal{\'a}n, Phys.~Rev.~B {\bf 55},  2476  (1997).

\bibitem{ESA70}
L. Esaki and R. Tsu, IBM~J.~Res.~Develop. {\bf 14},  61  (1970).

\bibitem{LAI93}
B. Laikhtman and D. Miller, Phys.~Rev.~B {\bf 48},  5395  (1993).

\bibitem{WAC97}
A. Wacker,  in {\em Theory of transport properties of semiconductor
  nanostructures}, edited by E. Sch{\"o}ll (Chapman and Hall, London, 1997), in
  print (cond-mat/9701105).

\bibitem{KAZ72}
R.~F. Kazarinov and R.~A. Suris, Sov.~Phys.~Semicond. {\bf 6},  120  (1972),
  [Fiz.~Tekh.~Poluprov. {\bf 6}, 148 (1972)].

\bibitem{ZHE93}
L. Zheng and A.~H. MacDonald, Phys.~Rev.~B {\bf 47},  10619  (1993).

\bibitem{MUR95}
S.~Q. Murphy, J.~P. Eisenstein, L.~N. Pfeiffer, and K.~W. West, Phys.~Rev.~B
  {\bf 52},  14825  (1995).

\bibitem{TUR96}
N. Turner, J.~T. Nicholls, E.~H. Linfield, K.~M. Brown, G.~A.~C. Jones, and
  D.~A. Ritchie, Phys.~Rev.~B {\bf 54},  10614  (1996).

\bibitem{WAC97b}
A. Wacker and A.-P. Jauho, Physica~Scripta {\bf T69},  321  (1997).

\bibitem{GOL88}
A. Gold, J. Serre, and A. Ghazali, Phys.~Rev.~B {\bf 37},  4589  (1988).

\bibitem{SER89}
J. Serre, A. Ghazali, and A. Gold, Phys.~Rev.~B {\bf 39},  8499  (1989).

\bibitem{GUI93}
P.~S.~S. Guimaraes, B.~J. Keay, J.~P. Kaminski, S.~J. Allen, P.~F. Hopkins,
  A.~C. Gossard, L.~T. Florez, and J.~P. Harbison, Phys.~Rev.~Lett. {\bf 70},
  3792  (1993).

\bibitem{KEA95a}
B.~J. Keay, S.~J. Allen, J. Gal{\'a}n, J.~P. Kaminski, K.~L. Champman, A.~C.
  Gossard, U. Bhattacharya, and M.~J.~M. Rodwell, Phys.~Rev.~Lett. {\bf 75},
  4098  (1995).

\bibitem{INA96}
J. I{\~n}arrea and G. Platero, Europhys.~Lett. {\bf 34},  43  (1996).

\bibitem{KEA95b}
B.~J. Keay, S. Zeuner, S.~J. Allen, K.~D. Maranowski, A.~C. Gossard, U.
  Bhattacharya, and M.~J.~M. Rodwell, Phys.~Rev.~Lett. {\bf 75},  4102  (1995).

\bibitem{ZEU96}
S. Zeuner, B.~J. Keay, S.~J. Allen, K.~D. Maranowski, A.~C. Gossard, U.
  Bhattacharya, and M.~J.~W. Rodwell, Phys.~Rev.~B {\bf 53},  1717  (1996).

\bibitem{PLA97}
G. Platero and R. Aguado, Appl.~Phys.~Lett. {\bf 70},  3546  (1997).

\bibitem{TIE63}
P.~K. Tien and J.~P. Gordon, Phys.~Rev. {\bf 129},  647  (1963).

\bibitem{TUC85}
J.~R. Tucker and M.~J. Feldman, Rev.~Mod.~Phys. {\bf 57},  1055  (1985).

\bibitem{WAG96}
M. Wagner, Phys.~Rev.~Lett. {\bf 76},  4010  (1996).

\bibitem{KOH59}
W. Kohn, Phys.~Rev. {\bf 115},  809  (1959).

\bibitem{MAH90}
G.~D. Mahan, {\em Many-Particle Physics} (Plenum, New York, 1990).

\bibitem{MIL94}
D. Miller and B. Laikhtman, Phys.~Rev.~B {\bf 50},  18426  (1994).

\bibitem{BRO90a}
G. Brozak, M. Helm, F. DeRosa, C.~H. Perry, M. Koza, R. Bhat, and S.~J. Allen,
  Phys.~Rev.~Lett. {\bf 64},  3163  (1990).

\bibitem{LEB70}
P.~A. Lebwohl and R. Tsu, J.~Appl.~Phys. {\bf 41},  2664  (1970).

\bibitem{SHI75}
A.~Y. Shik, Sov.~Phys.~Semicond. {\bf 8},  1195  (1975),
  [Fiz.~Tekh.~Poluprov.{\bf 8}, 1841 (1974)].

\bibitem{ADA93}
{\em Properties of Aluminium Gallium Arsenide}, edited by S. Adachi (INSPEC,
  London, 1993).

\bibitem{HEL90}
P. Helgesen and T.~G. Finstad,  in {\em Proceedings of the $14^{th}$ Nordic
  Semiconductor Meeting}, edited by O. Hansen (University of {\AA}rhus,
  {\AA}rhus, 1990), p.\ 323.

\bibitem{HEL91}
P. Helgesen, T.~G. Finstad, and K. Johannessen, J.~Appl.~Phys. {\bf 69},  2689
  (1991).

\bibitem{HIL96}
W. Hilber, M. Helm, K. Alavi, and R.~N. Pathak, Appl.~Phys.~Lett. {\bf 69},
  2528  (1996).

\bibitem{IGN95}
A.~A. Ignatov, E. Schomburg, J. Grenzer, K.~F. Renk, and E.~P. Dodin,
  Z.~Phys.~B {\bf 98},  187  (1995).

\bibitem{WAC97a}
A. Wacker, M. Moscoso, M. Kindelan, and L.~L. Bonilla, Phys.~Rev.~B {\bf 55},
  2466  (1997).

\bibitem{PRE94}
F. Prengel, A. Wacker, and E. Sch{\"o}ll, Phys.~Rev.~B {\bf 50},  1705  (1994),
  ibid {\bf 52}, 11518 (1995).

\bibitem{BON94}
L.~L. Bonilla, J. Gal{\'a}n, J.~A. Cuesta, F.~C. Mart\'{\i}nez, and J.~M.
  Molera, Phys.~Rev.~B {\bf 50},  8644  (1994).

\bibitem{WACp}
A. Wacker and A.-P. Jauho, phys.~status solidi (b)  , in print.

\bibitem{SZO92}
W. Szott, C. Jedrzejek, and W.~P. Kirk, Phys.~Rev.~B {\bf 45},  3565  (1992).

\bibitem{SHK84}
B.~I. Shklovskii and A.~L. Efros, {\em Electronic Properties of Doped
  Semiconductors} (Springer, Berlin, 1984).

\bibitem{STE67a}
F. Stern, Phys.~Rev.~Lett. {\bf 18},  546  (1967).

\bibitem{AND82a}
T. Ando, J.~Phys.~Soc.~Jpn. {\bf 51},  3215  (1982).

\bibitem{DAS83}
S. {Das Sarma}, Phys.~Rev.~Lett. {\bf 50},  211  (1983).

\bibitem{HU93}
B.~Y. Hu and S. {Das Sarma}, Phys.~Rev.~B {\bf 48},  5469  (1993).

\end{thebibliography}

\newpage

\newpage

\newpage

\newpage

\end{multicols}
\end{document}